# Learning to Prompt in the Classroom to Understand AI Limits: A pilot study

Emily Theophilou[b][0000-0001-8290-9944], Cansu Koyutürk[a], Mona Yavari[a], Sathya Bursic[a][0000-0001-8327-5007], Gregor Donabauer[a,c], Alessia Telari[a], Alessia Testa[a], Raffaele Boiano[d][0000-0002-3676-4163], Davinia Hernandez-Leo[b*][0000-0003-0548-7455], Martin Ruskov[e**][0000-0001-5337-0636], Davide Taibi[f***][0000-0003-0548-7455], Alessandro Gabbiadini[a][0000-0002-7593-8007], and Dimitri Ognibene[†a,g][0000-0002-9454-680X]

[a] Dept. Psychology, Università degli Studi di Milano Bicocca, Milan, Italy
[b] Universitat Pompeu Fabra, Barcelona, Spain
[c] University of Regensburg, Regensburg, Germany
[d] Politecnico di Milano, Milan, Italy
[e] Università degli Studi di Milano, Milan, Italy
[f] Istituto per le Tecnologie Didattiche (ITD-CNR), Palermo, Italy
[g] Università of Essex, Colchester, United Kingdom
*davinia.hernandez-leo@upf.edu, **martin.ruskov@unimi.it, ***davide.taibi@itd.cnr.it, ****alessandro.gabbiadini@unimib.it
†Lead PI, dimitri.ognibene@unimib.it

**Abstract.** Artificial intelligence's (AI) progress holds great promise in tackling pressing societal concerns such as health and climate. Large Language Models (LLM) and the derived chatbots, like ChatGPT, have highly improved the natural language processing capabilities of AI systems allowing them to process an unprecedented amount of unstructured data. However, the ensuing excitement has led to negative sentiments, even as AI methods demonstrate remarkable contributions (e.g. in health and genetics). A key factor contributing to this sentiment is the misleading perception that LLMs can effortlessly provide solutions across domains, ignoring their limitations such as hallucinations and reasoning constraints. Acknowledging AI fallibility is crucial to address the impact of dogmatic overconfidence in possibly erroneous suggestions generated by LLMs. At the same time, it can reduce fear and other negative attitudes toward AI. This necessitates comprehensive AI literacy interventions that educate the public about LLM constraints and effective usage techniques, i.e prompting strategies. With this aim, a pilot educational intervention was performed in a high school with 21 students. It involved presenting high-level concepts about intelligence, AI, and LLMs, followed by practical exercises involving ChatGPT in creating natural educational conversations and applying established prompting strategies. Encouraging preliminary results emerged, including high appreciation of the activity, improved interaction quality with the LLM, reduced negative AI sentiments, and a better grasp of limitations, specifically unreliability, limited understanding of commands leading to unsatisfactory responses, and limited presentation flexibility. Our aim is to explore AI acceptance factors and refine this approach for more controlled future studies.

# 1 Introduction

Artificial Intelligence (AI) technologies have gained significant prominence in contemporary society, permeating various facets of everyday life. AI is increasingly assuming a vital role in driving progress toward sustainable development worldwide in fields like healthcare, education, climate action [1, 2, 48, 52, 42, 58]. As an example, AI has already contributed to tackling medicine and health issues by improving diagnosis [9], developing new treatments [15, 25, 53], and supporting the overall care process at multiple scales. It also promises to help to deal with the chronic lack of expert personnel that is affecting many developing countries [62] both through training personnel and simplifying the medical procedures [52]. However, with all this potential comes big responsibility. While in the medical domain, the critical lack of personnel reduces the importance of the impact of the issue of jobs loss, several other problems must still be addressed. First, limited AI literacy may limit the gain for the countries where these tools would be more useful. A second issue is patient privacy, as the absence of a transparent and reliable process in place could lead to health data being used for unrelated applications of different entities, e.g. impacting patient access to job, insurance, and financial services [32]. Care must also be taken when applying AI decisions at multiple levels of the healthcare process as they may produce biassed results [41] resulting from biassed objectives and datasets. Moreover, determining the responsibilities in case of bad consequences of AI decisions is a complex topic that has been discussed for decades [67, 37].

With the magnitude of the contrasting positive and negative potential outcomes combined to the astonishing speed and complexity of the AI field, it was to be expected the rise of highly contrasting attitudes toward AI, extending from enthusiasm to phobia. Despite positive outcomes of AI systems, the recent advancements in AI have also sparked fears, anxiety, and negative attitudes particularly when machines begin to perform mindful tasks traditionally associated with humans [13, 22].

Media representations have often amplified these concerns by emphasising the negative consequences of AI and frequently depicting scenarios involving killer robots [30]. Such portrayals contribute to the magnification of AI anxiety. The impact of this negative sentiment toward AI can be dramatic, hindering trust and the acceptance and adoption of AI technologies and blocking the contributions they can provide. For instance, while AI diagnosis performance reaches or surpasses those of expert physicians, it will provide a real clinical benefit only if physicians will take into account its predictions [14]. Thus enabling healthcare professionals to achieve the right balance between trust and suspicion is crucial for achieving the full AI potential in medicine [14, 57]. The same balance is crucial to not miss the important opportunities that AI can provide in many domains and that anxiety-driven rejection or bans would hinder [23, 35, 40]. Understanding the causes of this anxiety is crucial for addressing these concerns. [24] identified three primary factors contributing to AI anxiety: (i) an over-emphasis on AI programs without considering the involvement of humans, (ii) confusion regarding the autonomy of computational entities and humans, and (iii) a flawed understanding of technological development. Addressing these factors through targeted literacy interventions is crucial in alleviating public concerns regarding AI advancements. Positive experiences with AI [43] and an understanding of how they work can shape positive attitudes towards AI [50] promoting its usage and acceptance among the public [35, 20]. Moreover, by delving into the inner workings of AI, individuals can develop critical perceptions toward these technologies [54] and become empowered to confidently embrace them.

### 1.1 The case of Large Language Models

The recent introduction of Large Language Models (LLM) like ChatGPT to the public may have been the tipping point for exasperating AI attitudes [18, 21]. LLM are machine learning models with a high number of parameters (from hundreds of millions for early models like BERT to hundreds of billions for GPT4) which are pre-trained to create lossy compression of large datasets through simple tasks, e.g. complete a statement or predict the next word, and can perform a variety of domain-independent tasks with little or no specific training and data [66, 59, 40]. LLM functioning is widely different from cognitive processes in biological brains and several LLM limits and vulnerabilities keep emerging [65, 56, 34, 4, 61, 3].

In particular, the tendency to make up responses to factual questions when they are not able to respond [12, 1]. Notwithstanding these limitations, the linguistic capabilities of LLM and ChatGPT have led to the strongest reactions comprising a letter signed by a number of experts calling for a stop of development of large models [2]. However, this call has been considered impractical or even counterproductive for democratic governance of these tools [23,40] and was not followed even by some of its main authors [39]. However, it still added fuel to the fire of AI phobia and anxiety.

Despite the growing familiarity with ChatGPT and its capabilities, there remains a lingering apprehension about the potential dominance of AI in various aspects of society. Some initial concerns have also emerged regarding its potential impact on educational aspects [17]. Educators, policymakers, and researchers are increasingly voicing concerns about the use of generative AI systems like ChatGPT in educational settings. One major concern revolves around the ethical considerations related to the use of generative AI systems by students [45]. Unethical practices, like using AI-generated content without appropriate attribution or engaging in plagiarism, pose challenges to academic integrity and raise questions about the responsibilities of both students and educators in the AI era. However, excluding ChatGPT from the classroom is not a viable solution, as its inclusion presents a valuable opportunity to familiarise students with the capabilities and limitations of generative AI tools [38, 60]. By explicitly incorporating ChatGPT into classroom activities, educators can provide students with insights and strategies for its proper utilisation, enabling them to effectively utilise this technology within a controlled and educational environment.

Students have a positive view of using ChatGPT as an educational tool, valuing its capabilities and finding it helpful for study and work. While acknowledging its potential for learning, students recognize the need for improvements and are mindful of its limitations [49]. The utilisation of ChatGPT in the classroom opens up opportunities for interactive and engaging learning experiences and prepares students for an increasingly AI-driven world. ChatGPT's capabilities in the classroom extend far beyond merely familiarising students with AI, as it demonstrates remarkable proficiency in covering diverse learning materials, spanning from coding [46] and microbiology [7] to media-related topics [44]. However, an essential aspect of utilising the full potential of ChatGPT lies in employing effective prompting strategies [64]. Carefully crafted prompts can guide ChatGPT's responses, leading to more accurate and informative outputs. This approach allows educators to align the AI system's responses with specific learning objectives, resulting in more targeted and meaningful interactions [29].

An important target for AI literacy, involving LLM, is defusing the rising and misleading feeling of being able to access and process any form of knowledge to solve problems in any domain with no effort or previous expertise in AI or problem domain. This widespread phenomenon stems from the lack of literacy on the inherent limitations of current LLMs, such as hallucinations, limited understanding, and reasoning constraints [12, 1]. By disregarding the boundaries of LLMs, individuals may fail to recognize the potential risks and inaccuracies that can arise from relying solely on their outputs. The recent widespread acceptance of generative AI LLM tools such as ChatGPT, highlights the necessity for informative interventions that educate users about realistic and comprehensive understandings of LLMs' capabilities and limitations. Such interventions can encourage users to exercise critical thinking when interpreting and applying knowledge generated by these models. Educators and researchers have been actively exploring and implementing diverse approaches to raise awareness and promote AI literacy within school environments [26, 51]. Recognizing the importance of going beyond theoretical aspects, these efforts aim to provide students with opportunities to expand their learning through hands-on experiences by incorporating practical activities, projects, and real-world applications of AI [26, 31].

As we embrace the new era of accessible AI tools, there is a noticeable lack of research on AI literacy interventions utilising ChatGPT. To address this gap and build upon existing concerns, this study aims to develop and evaluate an intervention focused on AI literacy, providing hands-on experience with ChatGPT. The primary goal is to assess the impact of this intervention on adolescents, exposing them to non-trivial tasks with ChatGPT to demonstrate its limitations while mitigating fears and negative attitudes towards AI. By engaging participants directly with the ChatGPT interface, the intervention aims to foster a deeper and more critical understanding of the technology and its potential limitations. This study specifically focuses on introducing adolescents to the strategy of prompting and examines their perceptions, emotions, interaction evaluations, and opinions toward ChatGPT. By evaluating the effectiveness of this educational approach, the study aims to offer valuable insights into reducing fear and promoting positive attitudes towards AI as well as introducing highly needed educational activities for the classroom about the novel concepts of prompting and LLMs.

## 2 Methodology

### 2.1 Participants and study design

A pilot study was conducted at a high school in Palermo, Sicily with a sample size of 21 students (n = 21; 33.3% male, 66.7% female; Ages 16 to 18, mean age = 16.3, SD = 0.57). The study was conducted within a formal school setting with students participating in a two hour-long AI workshop. Prior to the study students were informed about the research objectives and the purpose of the workshop and were asked to sign an electronic form to provide their consent to their participation in the study.

### 2.2 Learning design and study procedure

The pilot study was conducted as an informative educational workshop on AI. The aim of the workshop was to introduce students to the topic of AI and encourage them to explore and question the capabilities and limitations of ChatGPT. The study procedure, depicted in Figure 1, was designed to facilitate learning through active exploration. In particular, the educational learning plan saw two phases, the first one introduced stu-dent to AI and allowed them to freely explore the capabilities and limitations of ChatGPT and the second phase introduced students to prompting techniques to enhance ChatGPT's capabilities.

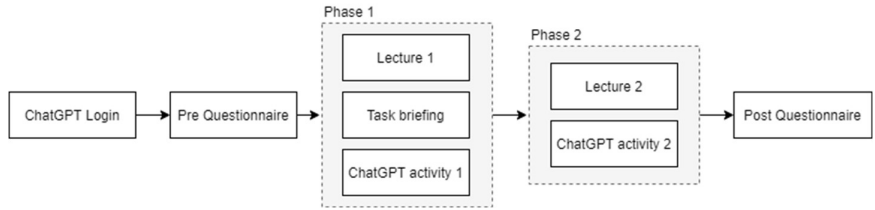

**Fig. 1.** Study design and educational learning plan.

The study procedure consisted of several key steps. Firstly, to minimize technical incidents influencing the results, students were instructed to access the ChatGPT page before they accessed the pre-questionnaire. After the completion of the pre-questionnaire, the instructor proceeded to deliver a presentation to introduce participants to topics related to AI applications, LLMs, and human intelligence vs artificial intelligence. Students were then provided with instructions for an activity that involved utilising ChatGPT. During this first activity students were asked to instruct ChatGPT to act as a personal teacher to educate in regards to the fundamental concepts of democracy (prior to the study the students had discussed the topic in class due to the Italian National Day and Republic Day, an initial internal trial was also conducted to verify ChatGPT's outputs for accuracy from an educational standpoint). The set of instructions included a variety of key points that the students should have as an outcome of their interaction. A few of the highlighted key points: ChatGPT should interactively explain the main concepts of democracy in a natural and not boring way. It should avoid long bulleted lists and alternate brief explanations with questions addressed to the user. Moreover, students were provided with a set of educational objectives the ChatGPT interaction would eventually generate.

The first ChatGPT activity lasted approximately 20 minutes and aimed to give first-hand experience to students in regard to the limited ability of ChatGPT to follow complex instructions. At its completion, the instructor proceeded to elaborate the limits of ChatGPT and then introduced the concept of prompting, providing a few simple examples. After receiving this information, students were given a second opportunity to instruct ChatGPT to act as a personal teacher. The task briefing was the same as in the first activity. At the end of the activity students accessed the post-questionnaire where they could also upload their interaction with ChatGPT.

### 2.3 Measures

*Perceived level of realistic and identity threat.* To measure the perceived level of realistic and identity threat generated by ChatGPT, a set of questions was adapted from the study of [63]. The questions were adapted to AI conversational skills and included items such as "In the long term, artificial intelligence is a direct threat to man's well-being and safety" and "Recent progress in artificial intelligence is challenging the true essence of what it means to be a human being". These sets of questions were part of both the pre and post questionnaires and were rated on a 7-point scale, with responses ranging from Strongly disagree (1) to Strongly agree (7).

*Self-Reported Emotions after interaction.* We proceeded to measure participants' emotions after their interaction with ChatGPT using the "The Discrete Emotions Questionnaire" adapted from [16]. Participants were then asked to report the degree of emotions they felt after the interaction with ChatGPT (anger, fear, disgust, anxiety, sadness, desire, happiness, joy) The items were anchored with (1) not at all to (7) very much.

*Interaction quality evaluation (UX).* Additionally, in the post-questionnaire, we proceed to collect data in regard to the interaction quality. In particular the subscales of "Semantic Differential Pragmatic dimension", "Semantic Differential Hedonic dimension", "Semantic Differential Human likeness", and "Social presence" were used from [19].

*Functionality of ChatGPT.* Moreover, in the post-questionnaire, a set of measures focused on evaluating students' perception of ChatGPTs functionality was included. Items were included to measure: (a) effort perceived to achieve desired ChatGPT behaviour, after their initial interaction with the AI tool, (b) perceived interaction improvement, after being introduced to prompting and engaging to a second interaction with the tool, and (c) ChatGPT capabilities.

*Open-ended question.* Lastly, to collect students' opinions in regard to the interaction with ChatGPT, the post-questionnaire included three open-ended questions to collect students opinions and thoughts in regards to; (a) positive aspects of the interaction, (b) negative aspects of the interaction, and (c) any additional noteworthy thoughts they wished to share.

Besides the aforementioned measures, we collected students' demographic data, their previous experiences with AI and ChatGPT, and in the post-questionnaire students were requested to paste their ChatGPT chat history.

### 2.4 Data analysis

To code and categorize the responses to open questions provided by participants, we used a classical social cognition model, the Stereotype Content Model (SCM), devised to describe the process of impression formation of social actors and groups, traditionally of human beings [10,11]. According to this theory, humans form and update their impression of others based on two fundamental dimensions: warmth, which involves characteristics such as friendliness, kindness, and trustworthiness – and competence – the ability to reach one's goals effectively. In the last decade, this model was applied to non-human agents like animals [47], brands [27], but also robots [6], chatbots [28], and artificial intelligence [36], showing promising results. In previous studies where people adopted warmth and competence to describe their AI interaction partner, they tend to express more competence-related judgments, and evaluate these agents as more competent than warm [36]. This may also depend on the particular AI system.

In this study, we decided to adopt this approach which summarizes social perception in two main dimensions. Some students' answers, though, were not targeting the perception of the chatbot per se but the whole educational activity and interaction with the composed system, referring to issues like creating an account or the excitement for their first interaction with an AI. Consequently, we devised a third category named "system" aimed at grouping these divergent records.

During the data collection process, an attention check was incorporated into each set of questions. As a result, the number of participants varied across questionnaires. The number of valid participants per questionnaire passing the test is reported in the results section.

## 3    Results

*Perceived level of realistic threat.* To create a composite measure for realistic threat, all five items on the scale were averaged together similar to previous work [63, 13]. Using this measure a dependent t-test revealed significant differences ($p<0.05$) between the pre (mPre= 4.17, SD=1.39) and post (mPost=3.73, SD=1.42) questionnaires (Fig 2). This suggests that participants' (n=20) realistic threat caused by AI decreased after the intervention. A closer look into the individual items, saw a significant decrease in participants' belief that AI is causing work loss for men (mPre= 4.6, SD=1.05, mPost=3.3, SD=1.49, $p<0.05$). However, participants' belief that AI will not replace workers from their duties remained unchanged after the intervention (mPre= 3.46, SD=1.25, mPost=3.46, SD=1.68, $p>0.05$). The remaining items saw a non-significant decrease after the intervention.

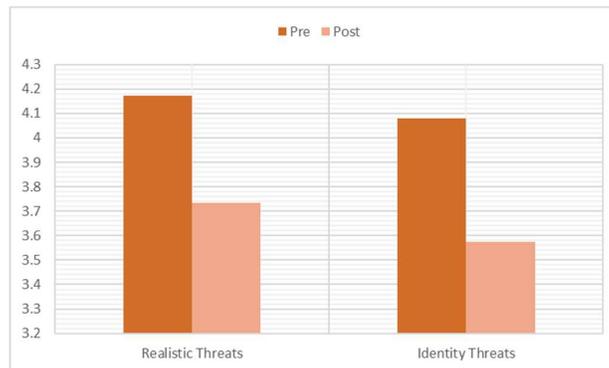

**Fig. 2.** Perceived level of realistic and identity threat aggregated average values before and after the intervention.

*Perceived level of identity threat.* A composite measure was created for identity threat by averaging all five items from the scale similar to previous work [63, 13]. A dependent t-test revealed a significant difference ($p<0.05$) between the pre (mPre= 4.08, SD=1.39) and post (mPost=3.57, SD=1.54) questionnaires (Fig 2). This finding indicates that participants' (n=20) AI identity threat significantly decreased after the intervention. A closer look into the individual items, saw a significant decrease in participants' belief that boundaries between man and machine are becoming less clear (mPre= 4.6, SD=1.29, mPost=3.73, SD=1.48, $p<0.05$). Despite improvements observed in the post-questionnaire, no statistically significant differences were identified among the remaining items of the scale.

*Self-Reported Emotions after interaction.* Participants (n=21) exhibited significantly higher positive emotions after their interaction with ChatGPT (mPositive= 3.48, SD=1.79, mNegative= 1.35, SD=0.91, $p<0.05$). Higher negative emotion was Anger (m=1.55 ,SD=1.43) whilst higher positive emotion was Serenity (m=3.65 ,SD=1.63). Lowest negative emotion was Sadness (m=1.2 ,SD=0.52) and lowest positive emotions were both Desire (m=3.4 ,SD=1.98) and Joy (m=3.4 ,SD=1.81). Values for all emotions are depicted in Fig 3.

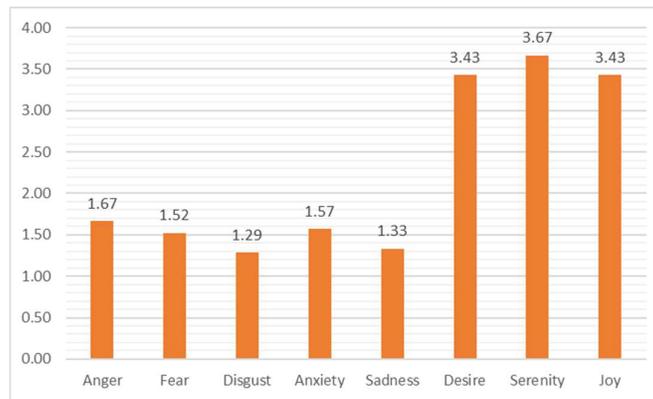

**Fig. 3.** Average values of self-reported emotions after the interaction with ChatGPT.

*Interaction quality evaluation (UX).* Under the first subscale "Perception of human likeness" students (n=20) perceived the interaction with ChatGPT more as an interaction with a machine rather than a human (m=2.9, SD=1.51), unnatural (m=3.6, SD=1.87), and artificial (m=3.1, SD=1.95). In the second subscale "Social Presence" the participants gave a substantially below average evaluation to the social aspects of the interaction (m=3.5, SD=1.67). With the highest rated item being that the chatbot was efficient in responding to the activities (m=4.6, SD=1.49) and the lowest rated item being that the chatbot engaged in a common task with them (m=2.9, SD=1.47). In the third subscale "Semantic Differential Hedonic dimension" participants overall found the experience enjoyable (m=4.75, SD=1.37) with the adjectives "Elegant, Good Quality, New, Created connections, Innovative, Presentable, and Engaging" receiving higher rating than their negative counter adjectives. In the final subscale "Semantic Differential Pragmatic dimension" the interaction was found predictable (m=4.85, SD=1.52) and manageable (m=5.52, SD=1.53).

*Functionality of ChatGPT.* In the first subscale, "Effort perceived to achieve desired ChatGPT behaviour" students' (n=21) responses indicated a neutral stance, with no strong agreement or disagreement on average. Participants reported that achieving the desired behaviour from ChatGPT required little effort (mean = 3.78, SD = 1.62), somewhat many attempts (mean = 4.47, SD = 1.61), somewhat more attempts were needed to refine the request (mean = 4.63, SD = 1.77), and the desired behaviour required increased understanding of how ChatGPT works (mean = 4.63, SD = 1.53).

After being introduced to the prompting strategies and completing the second activity students were asked to compare the two interactions. Compared to the first attempt, students found the results of the second interaction to be better (mean = 2.45, SD = 1.19), slightly more natural (mean = 4.55, SD = 1.61), and clearer (mean = 5.25, SD = 1.11). However, there was no agreement if the interaction was passively repeating content or more interactive (mean = 4, SD = 1.83).

Finally, in regards to the subscale of "ChatGPT capabilities", participants found ChatGPT intelligent rather than confused (mean = 3, SD = 2.01), intuitive instead of unable to adapt to requests (mean = 3.17, SD = 1.99), understanding of their questions (mean = 3.39, SD = 1.77), knowing what they asked (mean = 3.28, SD = 1.99), adapting to their questions rather than repeating the same mistakes (mean = 3.67, SD = 1.76), however, they reported the interactions as reading from an encyclopaedia rather than communicating with a human (mean = 3.67, SD = 1.76).

*Open-ended question.* The dimension that was most widely covered in the open answers was competence, with the theme that emerged most strongly being that ChatGPT was responsive and provided answers. This theme was supported by the responses of 8 participants. The responses were characterised by terms such as "immediate," as emphasised by participants P06 and P13, "correct" (P09), "exhaustive" (P04), and "interesting" (P10). A student also observed that the system was able to provide summarisations on request (P06). Another theme that emerged within this dimension with the support of 5 students is the system's usefulness. A student commented that this use of ChatGPT "could be useful for practicality and timing" (P20).

Negative aspects of competence that students commented on were repetitiveness, both in terms of them needing to repeat their questions and ChatGPT repeating responses. These were supported by the writing of 4 students each. The first one of these was mentioned with comments along the lines of "it started repeating the same things" (P06), and the second - along the lines of a student saying they "had to repeat several times to explain [themselves] again and more clearly the topics" (P11). One student also wrote that they had "to repeat [to the system to] to go slowly several times" (P12). Another theme of criticism, related to this need for repetition was supported by 5 students, and represented by writings stating that the system "did not answer as [the student] wanted to questions" (P17) and "the chat was purely notionistic" (P01). When it comes to the warmth dimension, only 3 students provided positive comments, giving a somewhat different spin to similar responses from the competence dimension. The main difference being that the focus in the responses is not on the system sharing its knowledge, but on it complying to students' requests. This is well represented by a student who wrote "that asking it to explain again in a clearer way, it acceded and fulfilled my requests" (P11). Criticisms that fall within the warmth dimension were about ChatGPT not being "natural" enough (P09) and not giving a sense of "a conversation with a human" (P20), "the little feeling in the replies" (P21), supported by 5 students. More precisely, they suggested that it should "briefly answer the questions that are asked" (P09) and it should not provide "answers that are taken from an encyclopedia" (P10), both of them also suggested it should be more human-like.

Finally, the third dimension that emerged was the perception of the system. Positive comments concerned the possibility to interact with artificial intelligence (supported by 3 students), and as a whole with a novel system (7 students). The two points were brought together by a student that expressed satisfaction of "dealing with a new reality, such as that of artificial intelligence" (P03). One student wrote to have found out "how much artificial intelligence can be useful in daily life[…] it helps to save time without being superficial in research" (P08). Others seconded that by writing that it "will surely be used for the future" (P17). In the majority of their negative comments regarding this dimension, students expressed views that the system needs to be improved, one saying that it's "still at an embryonic stage" (P02).

## 4   Discussion and conclusion

This study aimed to develop an AI literacy workshop using ChatGPT to enhance adolescents' understanding of AI limitations and mitigate fears and negative attitudes towards AI. The intervention successfully reduced adolescents' fears related to realistic and identity threats posed by AI advancements. The initial levels in the responses to the corresponding metrics demonstrated the presence of such a fear, similar to previous work [13]. Our study revealed that offering opportunities for guided non-trivial interactions with ChatGPT can effectively reduce the fear associated with AI advancements. In particular, a significant decrease was noticed in the items of fear of job loss and belief in the blurring boundaries between humans and machines. This positive shift in attitudes indicates that the exposure of adolescents to generative AI capabilities provided them a better understanding of how AI systems function and the impact they may have on various aspects of society, including the job market and human identity.

Regarding the overall experience, students rated the interaction with ChatGPT as enjoyable, eliciting positive emotions such as desire, serenity, and happiness. In some instances, students reported feelings of anger, which may be attributed to factors beyond the interaction with ChatGPT, such as their difficulties during the registration phase or the survey. This claim is further supported by comments students left in the open-ended responses.

In regard to evaluating the interaction of ChatGPT in terms of human likeness, students perceived ChatGPT as more of a machine than a human-like entity, describing it as unnatural and artificial [29]. This finding was persistent in the open-ended answers with students further describing their interactions with ChatGPT as repetitive. The social presence perceived during the interaction was limited with students reporting that the chatbot did not engage in common tasks with them. However, despite these perceptions, students found the experience enjoyable and manageable.

When comparing the two ChatGPT activities, the initial one without prompting strategies and the second one after being introduced to prompting, the students rated the second interaction with ChatGPT clearer, more natural and better than the initial attempt. A look into students' requests within ChatGPT, we observe more structured prompts as the interactions went on. Due to limitations in collecting the majority of students generated prompts, it was not feasible to derive more concrete results in regards to prompting skills improvement, however students reported perceived improvement in interaction with ChatGPT and understanding of its capabilities. Moreover, the incorporation of prompting strategies in the second ChatGPT activity had a profound impact on students' perceptions and evaluations of the overall interaction. Highlighting the importance of providing users with appropriate guidance and education to fully leverage the capabilities of AI systems [64].

Overall, the findings indicate that participants had a positive view of ChatGPT's capabilities, appreciating its intelligence, understanding, and adaptability similar to pre-vious work [49]. However, despite these positive evaluations of ChatGPT's capabilities, participants perceived the interactions as more akin to reading from an encyclopedia rather than engaging in human-like communication [29]. This suggests that while students recognized the intelligence and adaptability of ChatGPT, they also acknowledged a limitation in its ability to emulate human-like interactions. However, it is essential to consider that this perception may also be influenced by a possible misunderstanding of the question from the students' point of view. The novelty of interacting with ChatGPT might have led them to expect encyclopedic-style answers to their natural language questions.

Findings from the student open-ended responses provide further valuable insights into their experiences and perceptions of ChatGPT in three distinct dimensions: competence, warmth, and system perception. On a warmth level, it was considered low while acceding to help the user, however, this could be part of the alignment fine tuning applied to ChatGPT. Finally, the dimension of system perception received positive comments, centred around the excitement of interacting with AI. Students proceeded to share individual thoughts of how they believed that AI, as represented by ChatGPT, is likely to become increasingly valuable in various aspects of daily life and education. To our knowledge this is the first study that offers an exploration about the need and an approach to learn to prompt with LLMs in the classroom and how this facilitates reflection about AI limits. The qualitative answers to the open questions provide a deep understanding about the aspects under exploration, but they cannot be generalised. Even if in case studies in other contexts we expect similar conditions (e.g. limited current familiarity with ChatGPT), more studies will be needed to determine the generality of our findings.

As any study we report the following limitations. Our choice of interpretation model for the open questions followed from our data. The concise qualitative answers did not allow for a fine-grained classification like the one proposed by [19] that we adopted in our quantitative interaction quality evaluation.

In a naive parallel between these measures, the warmth dimension can efficiently capture the hedonic, social presence, and human-likeness dimensions, while the pragmatic quality dimension aligns with the competence aspect. However, it must be noted that the SCM was mostly designed with human actors and human-level linguistics [5] and functional cognitive capabilities in mind [34]. Instead, [19] propose measures that were initially applied to classical chatbots whose interaction capabilities were more restrictive, e.g. fixed agent-led instead of mixed-initiative dialog. Those chatbots were designed to effectively complete a specific task with a limited focus on natural and versatile interaction. Pragmatic value for these models usually refers to the complexity of the task and domain at hand, e.g. acquiring all the data necessary from the user and completing the operations requested. This measure may not directly map pragmatic linguistic skills [5], which were too limited in most old commercial chatbots. While later UX chatbot measures like those of [19] have been applied to more complex chatbots, they don't split clearly the perception of different types of linguistic [5] and emotional skills, which may affect items present in all four dimensions: pragmatic quality, hedonic quality, human-likeness, and social presence. SCM would instead collapse in the competence dimension both the semantic and pragmatic linguistic skills while the latter is domain independent and connected to the social domain. The disagreement between these measures were often reflected by disagreement between the annotators.

For example, the issues about repetitiveness of responses or need to repeat and reformulate a query were considered by the majority as lack of competence, thus following the selected SCM approach, a minority as social competence, following a line of reasoning more inline with the view of UX chatbot measure. This may explain the contrast between the interaction quality evaluation (that finds the system competent, and the open-ended answers analysis that presents several negative points on this aspect.

The UX chatbot and SCM measures may not be fully suited to cover for both versatility and fragility of modern LLM-based chatbots and in particular the interaction between their broader but fallible capabilities [34] as the tendency to diverge into hallucinations [55], especially during complex natural conversations [29], and the unnatural almost hardwired safeguard responses they present [33, 8]. To get a more detailed measure of users' perception of ChatGPT skills we added specific semantic differential measures "Functionality of ChatGPT" that non conclusively suggest a positive perception of ChatGPT's capabilities while being still limited in terms of natural interaction. In our future studies, we will extend the measures collected to account for these issues, for example adopting automatic tools for measurement of semantic and pragmatic precision [5].

The study was carried out as a field study within a school environment, but encountered certain challenges related to the accessibility of the ChatGPT website. Additionally, in some instances, students worked in pairs to complete the activity due to malfunctioning of some machines. While most students reported improving quality of the interaction during the activity only nine uploaded their in-class interaction due to technical issues. Only five out of nine interactions showed more than three attempts to improve the conversation modifying the prompts. Moreover, the number of questions in the survey may have tired the students and affected their answers. It is important to note that this was an exploratory pilot study with a relatively small sample size necessitates caution in generalising the findings.

In conclusion, our study suggests a significant impact of designing and developing AI literacy workshops with hands-on experience using ChatGPT. While with a limited number of participants, the intervention has shown to be an effective approach in enhancing participants' understanding of ChatGPT limitations and capabilities whilst also diminishing fears of identity and realistic threats caused by AI advancements. Lastly, the study successfully introduced participants to the effective use of prompting strategies, enhancing their interactions with ChatGPT.

To conclude, we highlight the need for novel measures of the linguistic aspects of user interaction with LLM based chatbots taking into account their non-transparent mechanisms and limitations as well as deal with large amounts of data [5]. In our future research in this line of inquiry we plan to replicate the study with a larger sample size, allowing for more comprehensive analyses and exploration for any correlations.

## Acknowledgements

This work has been partially funded by the Volkswagen Foundation (COURAGE project, no. 95567). TIDE-UPF also acknowledges the support by AEI/10.13039/501100011033 (PID2020-112584RB-C33, MDM-2015-0502) and by ICREA under the ICREA Academia programme (D. Hernández-Leo, Serra Hunter) and the Department of Research and Universities of the Government of Catalonia (SGR 00930). The authors thank Marco Marelli for the useful discussions on pragmatic linguistic skills.

# References

1. Bang, Y., Cahyawijaya, S., Lee, N., Dai, W., Su, D., Wilie, B., ... & Fung, P. (2023). A multitask, multilingual, multimodal evaluation of chatgpt on reasoning, hallucination, and interactivity. arXiv preprint arXiv:2302.04023.
2. Bengio Y., Russel S., Musk E., Wozniak S., & Harari Y.N. (2023). Pause Giant AI Experiments: An Open Letter. Future of Life Institute; https://futureoflife.org/open-letter/ pause-giant-ai-experiments/
3. Bishop, J. M. (2021). Artificial intelligence is stupid and causal reasoning will not fix it. Frontiers in Psychology, 11, 2603.
4. Borji, A. (2023). A categorical archive of chatgpt failures. arXiv preprint arXiv:2302.03494.
5. Bunt, H., & Petukhova, V. (2023). Semantic and pragmatic precision in conversational AI systems. Frontiers in Artificial Intelligence, 6, 896729.
6. Carpinella, C. M., Wyman, A. B., Perez, M. A., & Stroessner, S. J. (2017, March). The robotic social attributes scale (RoSAS) development and validation. In Proceedings of the 2017 ACM/IEEE International Conference on Human-Robot Interaction (pp. 254-262).
7. Das D, Kumar N, Longjam L, et al. (March 12, 2023) Assessing the Capability of ChatGPT in Answering First- and Second-Order Knowledge Questions on Microbiology as per Competency-Based Medical Education Curriculum. Cureus 15(3): e36034. DOI 10.7759/cureus.36034
8. Derner, E., & Batistič, K. (2023). Beyond the Safeguards: Exploring the Security Risks of ChatGPT. arXiv preprint arXiv:2305.08005.
9. Esteva, A., Kuprel, B., Novoa, R. A., Ko, J., Swetter, S. M., Blau, H. M., & Thrun, S. (2017). Dermatologist-level classification of skin cancer with deep neural networks. nature, 542(7639), 115-118.
10. Fiske, S. T., Cuddy, A. J. C., Glick, P., & Xu, J. (2002). A model of (often mixed) stereotype content: Competence and warmth respectively follow from perceived status and competition. Journal of Personality and Social Psychology, 82(6), 878–902.
11. Fiske, S. T., Xu, J., Cuddy, A. J. C., & Glick, P. (1999). (Dis)respecting versus (dis)liking: Status and interdependence predict ambivalent stereotypes of competence and warmth. Journal of Social Issues, 55(3), 473-489.
12. Floridi, L. (2023). AI as an agency without intelligence: on ChatGPT, large language models, and other generative models. Philosophy & Technology, 36(1), 15.
13. Gabbiadini, A., Ognibene, D., Baldissarri, C., & Manfredi, A. (2023). Does ChatGPT Pose a Threat to Human Identity? Available at SSRN: https://ssrn.com/abstract=4377900 or http://dx.doi.org/10.2139/ssrn.4377900.
14. Gaube, S., Suresh, H., Raue, M., Merritt, A., Berkowitz, S. J., Lermer, E., ... & Ghassemi, M. (2021). Do as AI say: susceptibility in deployment of clinical decision-aids. NPJ digital medicine, 4(1), 31.
15. Gupta, R., Srivastava, D., Sahu, M., Tiwari, S., Ambasta, R. K., & Kumar, P. (2021). Artificial intelligence to deep learning: machine intelligence approach for drug discovery. Molecular diversity, 25, 1315-1360.
16. Harmon-Jones, C., Bastian, B., & Harmon-Jones, E. (2016). The discrete emotions questionnaire: A new tool for measuring state self-reported emotions. PloS one, 11(8), e0159915.
17. Haque, M. U., Dharmadasa, I., Sworna, Z. T., Rajapakse, R. N., & Ahmad, H. (2022). " I think this is the most disruptive technology": Exploring Sentiments of ChatGPT Early Adopters using Twitter Data. arXiv preprint arXiv:2212.05856.
18. Harari, Y. N. (2018). Why technology favors tyranny. The Atlantic, 322(3), 64-73.

19. Haugeland, I. K. F., Følstad, A., Taylor, C., & Bjørkli, C. A. (2022). Understanding the user experience of customer service chatbots: An experimental study of chatbot interaction design. International Journal of Human-Computer Studies, 161, 102788.
20. Hyesun C., Prabu D. & Arun R. (2023) Trust in AI and Its Role in the Acceptance of AI Technologies, International Journal of Human–Computer Interaction, 39:9, 1727-1739, DOI: 10.1080/10447318.2022.2050543
21. Ipsos MORI, (2017). Public views of machine learning. Retrieved from https://royalsociety.org/~/media/policy/projects/machine-learning/publications/public-views-of-machine-learning-ipsos-mori.pdf, accessed 20 June 2019.
22. Dang,J. & Liu, L. (2021). Robots are friends as well as foes: Ambivalent attitudes toward mindful and mindless AI robots in the United States and China. Computers in Human Behavior, 115, 106612. ISSN 0747-5632. https://doi.org/10.1016/j.chb.2020.106612.
23. Ienca, M. (2023). Don't pause giant AI for the wrong reasons. Nature Machine Intelligence, 1-2.
24. Johnson, D.G., & Verdicchio, M. (2017). AI Anxiety. Journal of the Association for Information Science and Technology, 68, 2267-2270. https://doi.org/10.1002/asi.23867.
25. Jumper, J., Evans, R., Pritzel, A., Green, T., Figurnov, M., Ronneberger, O., ... & Hassabis, D. (2021). Highly accurate protein structure prediction with AlphaFold. Nature, 596(7873), 583-589.
26. Kandlhofer, M., Steinbauer, G., Hirschmugl-Gaisch, S. and Huber, P. (2016). Artificial intelligence and computer science in education: From kindergarten to university. In 2016 IEEE Frontiers in Education Conference (FIE), Erie, PA, USA (pp. 1-9). doi: 10.1109/FIE.2016.7757570.
27. Kervyn, N., Fiske, S. T., & Malone, C. (2012). Brands as intentional agents framework: How perceived intentions and ability can map brand perception. Journal of Consumer Psychology, 22(2), 166-176.
28. Khadpe, P., Krishna, R., Fei-Fei, L., Hancock, J. T., & Bernstein, M. S. (2020). Conceptual metaphors impact perceptions of human-AI collaboration. Proceedings of the ACM on Human-Computer Interaction, 4(CSCW2), 1-26
29. Koyutürk, C., Yavari, M., Theophilou, E., Bursic, S., Donabauer, G., Telari, A., ... & Ognibene, D. (2023). Developing Effective Educational Chatbots with ChatGPT prompts: Insights from Preliminary Tests in a Case Study on Social Media Literacy. arXiv preprint arXiv:2306.10645.
30. Lemay, David & Basnet, Ram & Doleck, Tenzin. (2020). Fearing the Robot Apocalypse: Correlates of AI Anxiety. International Journal of Learning Analytics and Artificial Intelligence for Education (iJAI). 2. 24. 10.3991/ijai.v2i2.16759.
31. Lomonaco, F., Taibi, D., Trianni, V., Buršić,S., Donabauer, G., and Ognibene, D. (2023). Yes, Echo-Chambers Mislead You Too: A Game-Based Educational Experience to Reveal the Impact of Social Media Personalization Algorithms. In G. Fulantelli, D. Burgos, G. Casalino, M. Cimitile, G. Lo Bosco, & D. Taibi (Eds.), Higher Education Learning Methodologies and Technologies Online: HELMeTO 2022 (pp. 26). Communications in Computer and Information Science, vol 1779. Springer, Cham. https://doi.org/10.1007/978-3-031-29800-4_26.
32. Luxton, D. D. (2014). Recommendations for the ethical use and design of artificial intelligent care providers. Artificial intelligence in medicine, 62(1), 1-10.
33. Brundage M., Mayer K., Eloundou T., Agarwal S., Adler S., Krueger G., Leike J.,and Mishkin P. .(2022, Mar) Lessons Learned on Language Model Safety and Misuse. [Online].Available: https://openai.com/research/language-model-safety-and-misuse
34. Mahowald, K., Ivanova, A. A., Blank, I. A., Kanwisher, N., Tenenbaum, J. B., & Fedorenko, E. (2023). Dissociating language and thought in large language models: a cognitive perspective. arXiv preprint arXiv:2301.06627.
35. Marangunić, N., & Granić, A. (2015). Technology acceptance model: A literature review from 1986 to 2013. Universal Access in the Information Society, 14(1), 81-95. https://doi.org/10.1007/s10209-014-0348-1.

36. McKee, K. R., Bai, X., & Fiske, S. (2021, February 26). Humans perceive warmth and competence in artificial intelligence. https://doi.org/10.31234/osf.io/5ursp
37. Mittelstadt, B. D., Allo, P., Taddeo, M., Wachter, S., & Floridi, L. (2016). The ethics of algorithms: Mapping the debate. Big Data & Society, 3(2), 2053951716679679.
38. Montanelli, S., Ruskov, M. (2023). A Systematic Literature Review of Online Collaborative Story Writing. In: Abdelnour Nocera, J., Kristín Lárusdóttir, M., Petrie, H., Piccinno, A., Winckler, M. (eds) Human-Computer Interaction – INTERACT 2023. INTERACT 2023. Lecture Notes in Computer Science, vol 14144. Springer, Cham. https://doi.org/10.1007/978-3-031-42286-7_5
39. Murgia, M., Bradshaw, T., Kinder, T. & Waters, R. (2023,APRIL 14). Elon Musk plans artificial intelligence start-up to rival OpenAI. Financial Times. https://www.ft.com/content/2a96995b-c799-4281-8b60-b235e84aefe4
40. Novelli, C., Casolari, F., Rotolo, A., Taddeo, M., & Floridi, L. (2023). Taking AI Risks Seriously: a Proposal for the AI Act. Available at SSRN 4447964.
41. Obermeyer, Z., Powers, B., Vogeli, C., & Mullainathan, S. (2019). Dissecting racial bias in an algorithm used to manage the health of populations. Science, 366(6464), 447-453.
42. Ognibene, D., Wilkens, R., Taibi, D., Hernández-Leo, D., Kruschwitz, U., Donabauer, G., ... & Eimler, S. (2023). Challenging social media threats using collective well-being-aware recommendation algorithms and an educational virtual companion. Frontiers in Artificial Intelligence, 5, 654930.
43. Oh, C., Song, J., Choi, J., Kim, S., Lee, S., & Suh, B. (2018, April). I lead, you help but only with enough details: Understanding user experience of co-creation with artificial intelligence. In Proceedings of the 2018 CHI Conference on Human Factors in Computing Systems (pp. 1-13). Montreal, QC, Canada. https://doi.org/10.1145/3173574.3174223
44. Pavlik, J. V. (2023). Collaborating With ChatGPT: Considering the Implications of Generative Artificial Intelligence for Journalism and Media Education. Journalism & Mass Communication Educator, 78(1), 84–93. https://doi.org/10.1177/10776958221149577
45. Qadir, J. (2023). Engineering Education in the Era of ChatGPT: Promise and Pitfalls of Generative AI for Education. In 2023 IEEE Global Engineering Education Conference (EDUCON), Kuwait, Kuwait (pp. 1-9). doi: 10.1109/EDUCON54358.2023.10125121.
46. Rahman M.M., Watanobe Y (2023). ChatGPT for Education and Research: Opportunities, Threats, and Strategies. Applied Sciences. 2023; 13(9):5783. https://doi.org/10.3390/app13095783
47. Sevillano, V., & Fiske, S. T. (2016). Warmth and competence in animals. Journal of Applied Social Psychology, 46(5), 276-293
48. Stahl, B. C. (2021). Artificial intelligence for a better future: an ecosystem perspective on the ethics of AI and emerging digital technologies (p. 124). Springer Nature.
49. Shoufan, A. (2023). Exploring Students' Perceptions of ChatGPT: Thematic Analysis and Follow-Up Survey. IEEE Access, 11, 38805–38818. https://doi.org/10.1109/ACCESS.2023.3268224
50. Shin, D. (2021). The effects of explainability and causability on perception, trust, and acceptance: Implications for explainable AI. International Journal of Human-Computer Studies, 146, 102551. ISSN 1071-5819. https://doi.org/10.1016/j.ijhcs.2020.102551.
51. Sánchez-Reina, J.R., Theophilou, E., Hernández-Leo, D., & Medina-Bravo, P. (2021). The power of beauty or the tyranny of algorithms: How do teens understand body image on Instagram? In B. Castillo-Abdul & V. García-Prieto (Eds.), Prosumidores emergentes: redes sociales, alfabetización y creación de contenidos (pp. 429-450). Sevilla: Editorial Dykinson S.L.
52. Sirmaçek, B., Gupta, S., Mallor, F., Azizpour, H., Ban, Y., Eivazi, H., ... & Vinuesa, R. (2023). The potential of artificial intelligence for achieving healthy and sustainable societies. In The Ethics of Artificial Intelligence for the Sustainable Development Goals (pp. 65-96). Cham: Springer International Publishing.

1653. Stokes, J. M., Yang, K., Swanson, K., Jin, W., Cubillos-Ruiz, A., Donghia, N. M., ... & Collins, J. J. (2020). A deep learning approach to antibiotic discovery. Cell, 180(4), 688-702.
54. Theophilou, E., Lomonaco, F., Donabauer, G., Ognibene, D., Sánchez-Reina, R.J., Hernàndez-Leo, D. (2023). AI and Narrative Scripts to Educate Adolescents About Social Media Algorithms: Insights About AI Overdependence, Trust and Awareness. In: Viberg, O., Jivet, I., Muñoz-Merino, P., Perifanou, M., Papathoma, T. (eds) Responsive and Sustainable Educational Futures. EC-TEL 2023. Lecture Notes in Computer Science, vol 14200. Springer, Cham. https://doi.org/10.1007/978-3-031-42682-7_28
55. Thorp, H. H. (2023). ChatGPT is fun, but not an author. Science, 379(6630), 313-313.
56. Valmeekam, K., Sreedharan, S., Marquez, M., Olmo, A., & Kambhampati, S. (2023). On the planning abilities of large language models (a critical investigation with a proposed benchmark). arXiv preprint arXiv:2302.06706.
57. Verghese, A., Shah, N. H., & Harrington, R. A. (2018). What this computer needs is a physician: humanism and artificial intelligence. Jama, 319(1), 19-20.
58. Wakunuma, K., Jiya, T., & Aliyu, S. (2020). Socio-ethical implications of using AI in accelerating SDG3 in Least Developed Countries. Journal of Responsible Technology, 4, 100006.
59. Wei, J., Bosma, M., Zhao, V. Y., Guu, K., Yu, A. W., Lester, B., ... & Le, Q. V. (2021). Finetuned language models are zero-shot learners. arXiv preprint arXiv:2109.01652
60. Woo, D. J., Guo, K., & Susanto, H. (2023). Cases of EFL Secondary Students' Prompt Engineering Pathways to Complete a Writing Task with ChatGPT. arXiv preprint arXiv:2307.05493.
61. Xu, L., Chen, Y., Cui, G., Gao, H., & Liu, Z. (2022). Exploring the universal vulnerability of prompt-based learning paradigm. arXiv preprint arXiv:2204.05239.
62. Yan, W., Qin, C., Tao, L., Guo, X., Liu, Q., Du, M., ... & Liu, J. (2023). Association between inequalities in human resources for health and all cause and cause specific mortality in 172 countries and territories, 1990-2019: observational study. bmj, 381.
63. Yogeeswaran, K., Złotowski, J., Livingstone, M., Bartneck, C., Sumioka, H., & Ishiguro, H. (2016). The interactive effects of robot anthropomorphism and robot ability on perceived threat and support for robotics research. Journal of Human-Robot Interaction, 5(2), 29-47.
64. Zamfirescu-Pereira, J. D., Wong, R. Y., Hartmann, B., & Yang, Q. (2023, April). Why Johnny can't prompt: how non-AI experts try (and fail) to design LLM prompts. In Proceedings of the 2023 CHI Conference on Human Factors in Computing Systems (pp. 1-21).
65. Zhang, H., Li, L. H., Meng, T., Chang, K. W., & Broeck, G. V. D. (2022). On the paradox of learning to reason from data. arXiv preprint arXiv:2205.11502.
66. Zhou, C., Li, Q., Li, C., Yu, J., Liu, Y., Wang, G., ... & Sun, L. (2023). A comprehensive survey on pretrained foundation models: A history from bert to chatgpt. arXiv preprint arXiv:2302.09419.
67. Ziosi, M., Mökander, J., Novelli, C., Casolari, F., Taddeo, M., & Floridi, L. (2023). The EU AI Liability Directive: shifting the burden from proof to evidence. AI & Society: Knowledge, Culture and Communication.